\documentclass[aps,pra,reprint,twocolumn,superscriptaddress,floatfix,nofootinbib,longbibliography]{revtex4-1}
\usepackage{graphicx,amsmath,amsfonts,amssymb,amsthm,xr}
\usepackage{epsfig,amsmath,amssymb,color,dsfont,upgreek,physics}
\usepackage{mathrsfs}
\usepackage{mathtools}
\usepackage{bbold}
\usepackage{adjustbox}
\usepackage{comment}
\usepackage{float}

\usepackage[bookmarks=true,colorlinks,linkcolor=OrangeRed,urlcolor=NavyBlue,citecolor=RoyalBlue]{hyperref}
\usepackage[dvipsnames]{xcolor}
\usepackage{orcidlink}

\definecolor{mygold}{rgb}{0.93,0.69,0.13}

\definecolor{mypurple}{rgb}{0.49,0.18,0.56}

\definecolor{mygreen}{rgb}{0,0.5,0}

\usepackage{braket}

\newcommand{\be}{\begin{equation}}
\newcommand{\ee}{\end{equation}}

\begin{document}

\title{
Tunable topological protection in Rydberg lattices via a novel quantum Monte Carlo approach 
}

\affiliation{
Theory of Quantum Matter Unit, Okinawa Institute of Science and
Technology Graduate University, Onna-son, Okinawa 904-0412, Japan
}
\affiliation{School of Physical and Chemical Sciences, Queen Mary University of London, London, E1 4NS, United Kingdom}
\affiliation{Max Planck Institute for the Physics of Complex Systems, N\"othnitzer Stra\ss e 38, 01187 Dresden, Germany}

\author{Pranay Patil${}^{\orcidlink{0000-0002-4554-6539}}$}
\email{pranay.patil15@gmail.com}
\affiliation{
Theory of Quantum Matter Unit, Okinawa Institute of Science and
Technology Graduate University, Onna-son, Okinawa 904-0412, Japan
}
\affiliation{Max Planck Institute for the Physics of Complex Systems, N\"othnitzer Stra\ss e 38, 01187 Dresden, Germany}

\author{Owen Benton}
\email{j.o.benton@qmul.ac.uk}
\affiliation{School of Physical and Chemical Sciences, Queen Mary University of London, London, E1 4NS, United Kingdom}
\affiliation{Max Planck Institute for the Physics of Complex Systems, N\"othnitzer Stra\ss e 38, 01187 Dresden, Germany}

\begin{abstract}
Rydberg atom arrays have recently been conjectured to host $Z_2$ quantum
spin liquids (QSLs) in certain parameter regimes. Due to the strong interactions
between these atoms, it is not possible to analytically study these systems,
and one must resort to Monte Carlo sampling of the path integral to reach
definite conclusions. We use a tailored update,
specifically designed to target the low energy excitations of the
QSL.
This allows us to reliably simulate Rydberg atoms on a triangular lattice
in the proposed QSL regime.
We identify a correlated paramagnetic phase at low temperatures
which hosts topological protection similar to a $Z_2$ spin liquid up to a
length scale tuned by Hamiltonian parameters. 
%
However, this correlated paramagnet seems to
be continuously connected to the trivial
paramagnetic regime and thus does not 
seem to be a true QSL.
%
This result indicates the feasibility
of Rydberg atom arrays to act as topological qubits.
\end{abstract}

\maketitle

Rydberg atom arrays have recently emerged as powerful platforms to simulate
quantum many-body systems
\cite{jaksch2000fast,lukin2001dipole,urban2009observation,browaeys2020many}.
The high degree of control generated by optical traps and laser detuning
allows the realization of simple models which may host entangled ground
states. One of the simplest models which approximate the behavior of such
arrays to a high degree is the PXP model
\cite{bluvstein2021controlling,pan2022composite,chen2021emergent},
which encodes the Coulomb repulsion between the large electron clouds of
neighboring excited atoms as a simple exclusion rule which prevents
co-existing excitations. 
It has been suggested
\cite{Vryd,samajdar2021quantum,yan2022triangular}
that atoms on triangular and Kagom\'e
lattices can realize a $Z_2$ quantum spin liquid.
Experiment evidence also supports this 
proposal \cite{semeghini2021probing}.
%
Such a state is not only of interest from a theoretical
perspective\cite{zhou2017quantum,savary2016quantum},
but also serves as the cornerstone for topologically protected quantum
computing\cite{kitaev2003fault}.

The strong correlations inherent to the
PXP model make it prohibitively
difficult to
understand the behavior of the system analytically. Thus, 
density matrix renormalization group (DMRG) simulations have been
used to show the presence of a QSL in a PXP model defined on the links of a  Kagom\'e lattice
\cite{Vryd}.
However, due to the nature of the method, these studies have been
limited to the ground state and to pseudo-1D geometries. 

Quantum
Monte Carlo (QMC) simulations have also been employed to simulate a
quantum dimer model with added terms to emulate the PXP
Hamiltonian, and these have shown that it is possible to achieve
QSL phases in the phase diagram of such a model\cite{yan2022triangular}
on the links of a triangular lattice.
However, the mapping from the PXP model to a quantum dimer model(QDM) is approximate, and it is not clear how applicable the results are to the original model. 
%
In this Article, we will address that issue by
applying QMC simulations to the PXP model on the triangular lattice.
To enable this, we have developed a novel Monte Carlo update which makes of intuition derived from QDM studies.
%
For QDMs with built-in resonating terms, a novel update
called the sweeping cluster algorithm has been used
\cite{yan2019sweeping,yan2022global}
to efficiently sample the configuration space in the corresponding
QMC simulation. As the PXP model does not explicitly have such
a resonance term, we have developed here an update which identifies the
resonances emergent at low temperature, and utilize a sampling which
toggles between various resonances, just as expected for a QSL.
Although there already exist local and cluster algorithms for
the PXP and similar models for Rydberg atoms
\cite{Melko,yue2021order,yan2019sweeping,biswas2016quantum,hearth2022quantum,wu2023u,wu2021z},
we explicitly show that the resonance update is crucial for
efficient sampling in the proposed QSL regime.
This algorithm can be easily generalized to other model Hamiltonians
which may host QSL physics, and is expected to provide a significant
improvement in sampling efficiency. We use our algorithm to show
the presence of a correlated paramagnet which resembles a QSL
at small scale in the PXP model on the 
triangular lattice,
but which seems to be smoothly connected to a trivial paramagnetic phase, even at zero temperature.

\begin{figure}[t]
\includegraphics[width=\hsize]{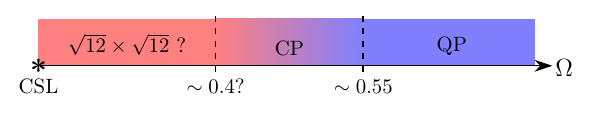}
\caption{
Zero temperature phase diagram for the $PXP$ model
on the triangular lattice as a function of the
strength of quantum fluctuations $\Omega$.
Strictly at $\Omega=0$, the ground state is a
$Z_2$ classical spin liquid (CSL) which has
short-ranged correlations. For small
$\Omega$, comparisons with the quantum dimer
model suggest the appearance of a $\sqrt{12}\times
\sqrt{12}$ ordering \cite{moessner2001resonating}. We have investigated the 
ground state for $\Omega>0.4$ and find a correlated
paramagnet (CP) which exhibits behavior consistent
with a $Z_2$ QSL up to intermediate system sizes,
followed by a simple quantum
paramagnet (QP) for large $\Omega$.
}
\label{FigQPT}
\end{figure}

The central numerical results which we are able to
achieve using the algorithm described above can be
summarized as follows : Fredenhagen-Marcu order
parameters \cite{FMorg,FP}, which are used to
identify topological ordering in dilute dimer models,
show a parameter regime which is consistent with
a $Z_2$ QSL. In addition, the dimer and energy
densities show changes in behavior
consistent with a transition
between two phases as a function of the control
parameter. However, these changes do not sharpen into
non-analyticities with increasing size (as expected
at a phase transition) and the energy density as
a function of temperature reveals a finite gap in
the large size limit, with no indication of 
a gap closing on the phase diagram. 
%
Since the simulations are carried out at temperatures substantially below the measured excitation gap, and since calculations of the entropy reveal that we reach the ground state regime, we do not believe that the smooth crossover from the correlated to trivial paramagnet is only a matter of finite temperature.
%
Rather, we conclude
that in the parameter regime we have been able to
analyze, there exists only a correlated paramagnetic
ground state (Fig.~\ref{FigQPT}), which behaves 
similarly to a $Z_2$ QSL
for intermediate system sizes
but is not a true QSL.

\begin{figure}[t]
\includegraphics[width=\hsize]{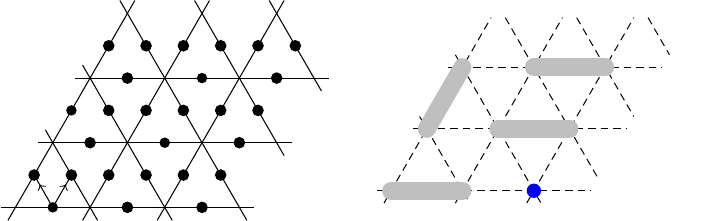}
\caption{
(Left) Rydberg atoms living on the links of a triangular lattice.
Each upward pointing triangle is a
unit cell, and the vectors within a
unit cell used to define the structure
factor (Eq.~\eqref{eqSF})
are shown for the first cell.
(Right) Excitations can be seen as dimers occupying the relevant links.
For a $3\times 3$ lattice, there is an odd number of lattice sites,
and thus there must be at least one site not covered by a dimer.
}
\label{FigTL}
\end{figure}

\section{Model and formulation of SSEQMC}\label{sSSE}

The most common model\cite{Vryd,yan2023emergent}
for Rydberg atom arrays is given by hard-core bosons
with a strong repulsive interaction and a kinetic term which
creates/annihilates bosons. In the limit of infinite repulsion,
the kinetic term can act only if every neighboring location is
empty. This reduces to the PXP model \cite{turner2018quantum},
where the P stands for a projection operator which enforces
zero occupancy on the required locations, and $X$ denotes a transverse
field in spin language ($b+b^{\dagger}$ in boson language).
We define our system on a triangular lattice with $L^2$ sites ($N=3L^2$ links),
where $L$ is the linear dimension of the system
(this can be seen as a 2D version of the model studied in
Ref.~\cite{fendley2004competing}).
The Rydberg atoms are placed at the link centers
as shown in Fig.~\ref{FigTL}.
For ease of comparison we begin with the paradigmatic
Hamiltonian for Rydberg atom arrays, parameterized 
in the same way as Ref.~\cite{Vryd} 
:
\be\label{Hryd}
H=\frac{\Omega}{2}\sum_i(b_i+b_i^{\dagger})-\delta\sum_i n_i+
\frac{1}{2}\sum_{i,j} V(|i-j|)n_in_j.
\ee
The Coulomb repulsion due to the Rydberg blockade is encoded in $V(|i-j|)$,
which is known to have the functional
form $V(r)\sim 1/r^6$ \cite{urban2009observation}.
As this is extremely short range, we make the approximation that for
nearest neighbors $V(r)\to\infty$, and $V(r)=0$ otherwise. This provides
us with the effective PXP model discussed above.
We have effectively removed all states from our boson occupation basis
which have bosons on nearest neighbors. Within this restricted space, the
Hamiltonian simply reduces to
\begin{equation}
H=\frac{\Omega}{2}
\sum_i(b_i+b_i^{\dagger})
- \delta\sum_i n_i.
\label{eq:H_pxp}
\end{equation}
It is important
to note here that even though it is not apparent from Eq. (\ref{eq:H_pxp}), the
restricted Hilbert space renders the system genuinely interacting.
As there is a freedom of one energy scale, we set $\delta=1$ for
all our numerical results, unless otherwise mentioned.

We can also express this as a spin model, using the mapping between hard-core bosons and spin-1/2 operators:
$
b_i^{\dagger} \to \sigma_i^+; \ b
_i \to \sigma_i^-;
n_i \to \frac{1}{2}(\sigma^z_i + 1).
$

In the spin language the Hamiltonian takes the form of
non-interacting spins in a global tilted field
(up to constant terms):
\begin{equation}
H=\frac{\Omega}{2}
\sum_i\sigma^x_i
- \frac{\delta}{2}\sum_i \sigma^z_i               
\label{eq:H_spin}
\end{equation}
However, it is worth noting here again that due to the
restricted Hilbert space allowed by the Rydberg constraint, the spins
should be considered to be strongly interacting, and thus capable
of interesting many-body physics.

In the restricted space, the action of the chemical potential is to maximize
the density of bosons, thus leading to fully packed configurations with
no neighboring locations simultaneously occupied by bosons. In contrast, the
kinetic term acts by creating resonances between the occupied and
unoccupied state, and in many cases leads to a phase continuously connected
to the trivial $x$-polarized paramagnet one would expect 
in the limit of large transverse field.

\begin{figure}[t]
\includegraphics[width=\hsize]{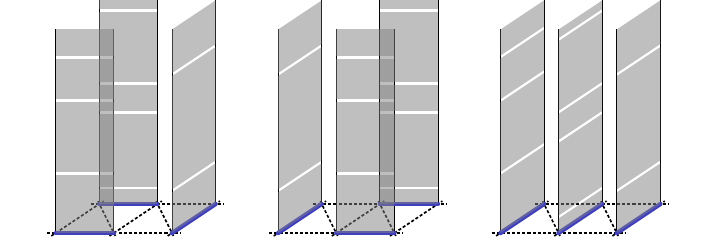}
\caption{
Path integral configurations in the classical spin liquid regimes.
Quantum fluctuations can be seen as breaks in the imaginary time
direction, and do not produce non-trivial structural changes.
}
\label{FigCSL}
\end{figure}

A spin liquid ground state may be realized by placing Rydberg atoms on the
links of the triangular lattice, as shown in Fig.~\ref{FigTL}.
The Rydberg constraint then excludes a simultaneous excitation of two
links which share a vertex. Although the constraint does not function exactly
in this manner for the native Rydberg models, it has recently been shown
that Rydberg ``gadgets'' can be built into the experimental set-up to
realize the dimer model accurately\cite{zeng2024quantum}.
This makes the system similar to a hard core
dimer model, where a Rydberg excitation can be understood as a dimer living
on the corresponding link, a Rydberg ground state as the absence of a dimer,
and the exclusion requirement now implies that
two dimers do not share an end point. In the classical limit $\Omega=0$,
the system now hosts a degenerate set of all fully packed dimer coverings
as its ground state, where the fully packed requirement comes from the
chemical potential term controlled by $\delta$. This suggests that for
$\Omega\ll\delta$, it may be possible to generate superpositions of these
dimer coverings, and thus a $Z_2$ spin liquid.

\subsection{SSEQMC to path integral configurations}

QMC involves sampling of path integral configurations which can be understood
from the Trotter decomposition of the partition function $Z=Tr[e^{-\beta H}]$.
Writing $e^{-\beta H}=(e^{-\Delta\tau H})^n$, and inserting a complete basis
after each $e^{-\Delta\tau H}$ leads to configurations defined by a collection
of basis states $\{\alpha_0,\alpha_1...\alpha_{n-1}\}$ (where $\alpha_i$
denotes a product state in our local basis), and this set is usually
stacked vertically in the third dimension to visualize a path integral
configuration.

\begin{figure}[t]
\includegraphics[width=\hsize]{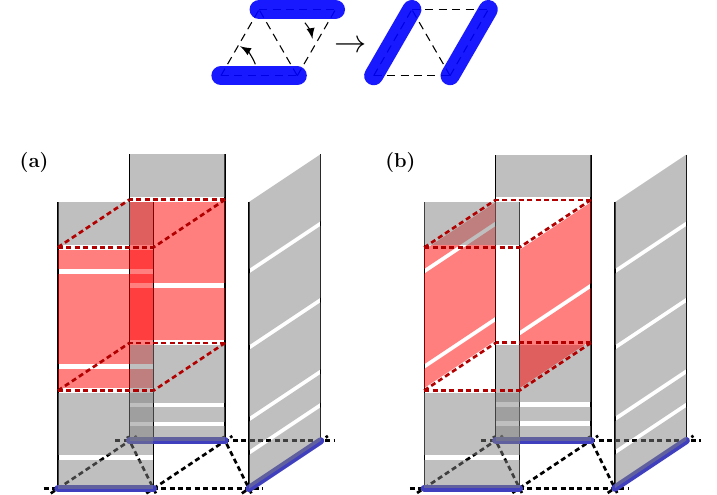}
\caption{
(Top) Two-dimer resonance on the triangular lattice.
(a) and (b) path integral configurations which contribute
to the partition function, related by the resonance shown above.
}
\label{FigQSL}
\end{figure}

For our simulations, we use quantum Monte Carlo simulations in the
stochastic series expansion formalism (SSEQMC) \cite{sandvik1992generalization}.
In our case, the first step in this formalism involves writing the
Hamiltonian as a sum of local terms $H=h_1+h_2+..+h_m$, which do
not have a branching property, i.e. $h_k\ket{i}=c^i_k\ket{j}$, where $\ket{i}$
and $\ket{j}$ are product states, and the set $\{c^i_k\}$
form the corresponding numerical coefficients. Thus each $h_k$ is a bijective
map from a product state to another product state in our chosen basis.
For the next step, we expand the partition function, $Z=Tr[e^{-\beta H}]$,
as $\sum_n\frac{(-\beta)^n}{n!}\sum_{s_n}Tr[h_{s^1_n}h_{s^2_n}...h_{s^n_n}]$,
where $s_n$ denotes a string of size $n$ and specifies the local terms
participating in the specific series of interest.
As a final step to help us visualize these strings as path integral
configurations, we insert complete basis sets in the boson
occupation basis (denoted by
$\sum_{\alpha}\ket{\alpha}\bra{\alpha}$)
between each pair $h_{s^m_n}h_{s^{m+1}_n}$ and for the trace. Due to the
non-branching property discussed above, only the sum over the basis set
for the trace remains. We denote this by
$\sum_{\alpha_0}\ket{\alpha_0}\bra{\alpha_0}$, and for a fixed $\alpha_0$
and particular $s_n$, all the intermediate basis states are specified as
$h_{s^1_n}\ket{\alpha_0}$ returns a unique basis state, $h_{s^2_n}$ acting
on that state again returns a unique basis state and so on.
We denote the intermediate states by
$\ket{\alpha_1},\ket{\alpha_2}..\ket{\alpha_{n-1}}$. Each state is
represented by a dimer covering on the triangular lattice, and we denote
the collection of states $\{\ket{\alpha_0},\ket{\alpha_1},\ket{\alpha_2}...\}$
as a path integral configuration by stacking them vertically in sequence.
This creates a three-dimensional configuration (as shown in Figs.~\ref{FigCSL}
and ~\ref{FigQSL});
identical configurations are also generated by carrying out
a vanilla Trotter decomposition of $Tr[e^{-\beta H}]$, and the vertical
dimension is usually referred to as the imaginary time direction.

\noindent
\section{Path integral configurations}\label{sPI}

To motivate the sampling algorithm which we use in our Monte Carlo simulation,
we discuss the possible path integral configurations which dominate
the partition function of classical and quantum spin liquids.


Let us begin now with the purely classical limit $\Omega=0$. As all
individual terms in the Hamiltonian
are now diagonal, all basis states in a path integral configuration must
be identical (as $\braket{a^1|H|a^2}=0$ if $\ket{a^1}\neq\ket{a^2}$).
Thus the configuration can be simply imagined as a single fully packed
dimer covering extended into the time dimension. The quantum fluctuation
which we introduce for non-zero $\Omega$ acts on a single link by either
destroying or creating a dimer. For
small $\Omega$, the effect of this on the path integral configurations can be
understood as small segments of dimer absence which appear as breaks in the
otherwise continuous dimer world line which we expect in the classical limit.
Cartoons for this type of configuration are shown in Fig.~\ref{FigCSL}.
The ensemble of path integral configurations still retains its classical
signature in this limit, and various configurations can be generated by
taking different dimer coverings, drawing them out in the time direction,
and populating them with the small breaks generated by the quantum
fluctuation. These configurations are expected in the $\Omega\ll\delta$
limit, and are shown in Fig.~\ref{FigCSL}.
The inverse temperature $\beta$ sets the length scale for the imaginary
time direction, and plays an important role here, as it controls how many
breaks are allowed in a particular configuration. Naively, the size and density
of the breaks would be expected to be independent of size of the system in
both space and time.

For the cartoon picture of a classical spin liquid
discussed here, the basis states in a particular path integral configuration
are more or less perfectly correlated. For a quantum spin liquid, this
would not be the case, as we should expect events in imaginary time which
resemble resonances between different dimer coverings. Examples of two and
three dimer resonances on the triangular lattice are shown in Fig.~\ref{FigQSL}.
In the path integral representation, these events can occur if there are
coordinated breaks on links which are compatible with a resonance. For
$\Omega\approx O(\delta)$, this is statistically likely, and this event
manifests in the path integral configuration as a sudden resonance with a
short span in imaginary time. Due to the trace condition, the resonance should
undo itself as a later point in imaginary time, thus leading to a pair of
resonances. A cartoon for such a configuration is shown in Fig.~\ref{FigQSL},
along with the configuration generated by flipping the column between the
resonances. The latter configuration resembles the classical spin liquid
configurations, and these are also present in the ensemble of the quantum
spin liquid. However, notice that the locations of the breaks in the classical
like configuration are correlated. This would not be the case for the classical
spin liquid, as the positions of quantum fluctuations are not expected to be
correlated in imaginary time.

Thus the signature of the path integral configurations of a quantum spin liquid
are these pairs of resonances which are well separated in imaginary time.
For small $\beta$, there is not sufficient volume in the time direction to
accommodate these pairs. This is the standard interpretation of the
effect of temperature on low energy quantum fluctuations, and leads to a
loss of such ``coherent'' resonances at intermediate temperatures.

\noindent
\section{Efficient Sampling}\label{sES}

We have understood the qualitative features of path integral configurations
of both classical and quantum spin liquids above. Now we can use these
to design efficient sampling algorithms.
We begin with the classical case. Updates local in time and local or non-local
in space, such as those introduced in
\cite{Melko,patil2023quantum},
can be used to efficiently sample fluctuations which are local
in time, such as the small segment breaks discussed above. However, these
updates will statistically never be able to sample different dimer coverings,
which control the large scale spatial structure of the configurations.
To generate new coverings, one can use a spatial worm update, which is
extended in the time direction. Practically, this corresponds to picking a
particular time slice and using its dimer covering to build a closed loop.
Along this closed loop, the status of the links can be toggled, leading to
a new and valid covering. Before this can be executed, one must build the
extension in the time direction. One simple way to do this is to simply ignore
all the breaks in the imaginary time direction, and consider the entire
world line of a dimer to be a single object. This intuition can be implemented
in a robust microscopic update which respects detailed balance in multiple
ways, and the version we have chosen is discussed below. A similar update has
recently been utilized to study the Kagom\'e geometry for the same model
\cite{wang2024renormalized}.
We find that our update is efficient at sampling in the classical spin
liquid regime.

\subsection{Classical worm update}

\begin{figure*}[t]
\includegraphics[width=\hsize]{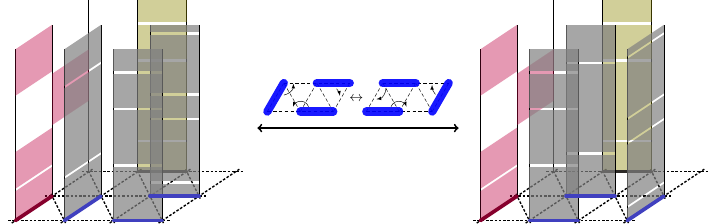}
\caption{
An example of a classical worm update for a three dimer resonance. The shaded
regions correspond to dimer occupation at the corresponding imaginary time
slice. The olive green
membrane is a satisfying dimer but does not happen to be a part of the loop
along which dimers are flipped, highlighting that not all satisfying dimers
need to participate in an update. The red membranes do not satisfy the classical
condition (due to a neighbor being occupied by a dimer for some region of
imaginary time).
}
\label{FigCW}
\end{figure*}

To carry out the classical worm update in our path integral picture, we focus
on links which are occupied by a dimer for the majority of the imaginary time
dimension, and have spatial
neighbors which are not occupied by a dimer on any
time slice. We refer to this as the classical condition.
An example of a configuration like this is shown in Fig.~\ref{FigCW}.
Once this is done, we consider all such links to be stitched completely in
imaginary time,
i.e. the entire temporal direction configuration of a particular site
moves as one unit. One can now use the standard worm update to carry
out a spatial loop or string move which moves the temporal patches
around. The process implemented by this update is shown in Fig.~\ref{FigCW}.
As we are considering dimer models which are not fully packed, the moves
generated by this update can often be open strings, and the detailed
balance for the end points must be carried out exactly in the manner as
described for the rod diffusion update in Ref.~\cite{patil2023quantum}.
As an added complication, the classical worm update
includes the step of identifying links which satisfy the classical condition
and we must ensure that this step also satisfies
balance. To see this, we can visualize the update as only considering the
set of sites satisfying the classical condition. The update then only moves
the relevant temporal patches 
in a manner which creates another legal configuration, as shown
in Fig.~\ref{FigCW}.
Links which were initially rejected due to one of neighbors being occupied
by a dimer in a non-zero region of imaginary time, are unaffected by this update,
as are their neighbors. Note that the neighbors are also rejected for the same
reason, i.e. its neighbor (the link discussed in the previous sentence) is
occupied by a dimer on at least one imaginary time slice.
Following this argument for all sites, we see that the number of links which
satisfy the classical condition before and after the update is exactly the
same. This implies that picking the link which will reverse the move out
of the set of satisfying links has the same probability as picking the first
link for the update in the forward direction. The probability that the loop
built is balanced is ensured by the rod diffusion update \cite{patil2023quantum}.

\subsection{Quantum resonance update}

Turning now to the quantum case, we begin by observing that the extended
worm update discussed above will not be able to produce the types of
configuration changes seen in Fig.~\ref{FigQSL}.
Thus these updates will completely miss the configurations which are a
signature of the quantum spin liquid. We remedy this by proposing an update
which specifically searches for possible pairs of resonances in the
path integral configuration. Schematically, this proceeds by picking a
random slice in imaginary time and first sweeping later positions in
imaginary time to identify a possible location where a resonance is possible.
Once this is found, we search in slices below the starting slice to see if
we can find the same resonance. If so, we have identified a column which
is bounded by a pair of identical resonances, and we can flip the dimers
in the resonance to get a new configuration, as shown in Fig.~\ref{FigQSL}.
The mechanism of maintaining detailed balance during this process is quite
involved and described in detail below.
In the following, we refer to this update as the quantum resonance.
We find that this update is essential in the correct sampling of path
integral configurations in parameter regimes where the ground state
resembles a $Z_2$ QSL.
The resonances discussed above have already been used as
an order parameter to detect a stable topological phase
for a perturbed toric code \cite{wu2012phase}.

We explicitly sample the quantum resonances expected in the imaginary time
direction using a construction of finite imaginary time tubes (with well defined
ends) whose spatial cross-section
resembles a dimer resonance on a closed loop. The update is carried out using
a fixed loop size (smallest size being four, which corresponds to a
nearest neighbor resonance as shown in Fig.~\ref{FigQSL})
in the following manner :\\

\noindent
1. We first calculate the maximal number of dimers on any single slice in the
imaginary time direction in the current path integral configuration. Next, we
identify all slices which have this maximal occupation, and pick one of these
slices at random with uniform probability. For later reference, we label this
slice at $s_{init}$. The column which we will flip will have end regions on
either side of $s_{init}$.\\

\noindent
2. We list all the dimers present in $s_{init}$ and store this as a reference
spatial configuration. Now we sequentially travel up in the imaginary time direction
and attempt to identify a suitable end region for our column. This is done at each
subsequent slice by defining a ``fuzzy'' region of size $N/\Omega$, as this is the
approximate size of imaginary time in which we can expect one dimer
creation/annihilation event per lattice link.\\

\begin{figure}[t]
\includegraphics[width=0.8\hsize]{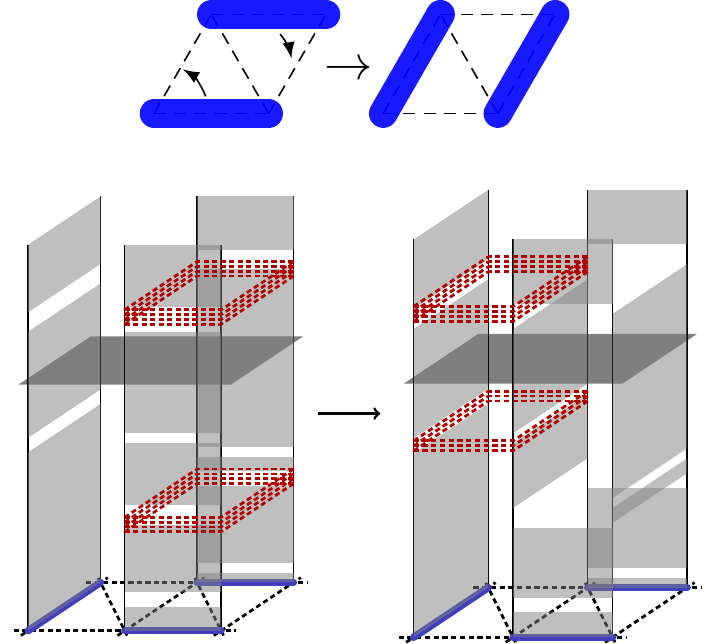}
\caption{
The quantum resonance update for the resonance shown (top). The update begins by using the
reference state at a randomly chosen slice $s_{init}$ (shown as a horizontal plane here),
and sweeping above and below this
slice to identify fuzzy regions (marked by dashed red lines) where both the dimers are
annihilated. These regions mark the ends of the update column, and we flip this column.
The new configuration generated in the particular
case shown does not generate the same modification on
the backward move, as the left plaquette has become
flippable now and is found before the right one
(when searching below the $s_{init}$ plane). Thus
this particular update does not obviously satisfy
detailed balance and is rejected (discussed in detail
in the sixth step of the quantum resonance update.)
}
\label{FigQW}
\end{figure}

\noindent
3. For a resonance of size $l$, the number of dimers which must be created/annihilated
in the fuzzy region must be $l/2$ (as shown for $l=4$ in Fig.~\ref{FigQW}). If this
condition is satisfied in the fuzzy region, we check if the identified list of $l/2$
dimers (label this list as $d_{l/2}$)
form a flippable loop. To simplify detailed balance, we consider the condition where
only one such loop exists.
This is technically challenging, and there are multiple
ways to carry out this identification. We use a simple method of looping all possible
neighboring links of $d_{l/2}$, and checking if this together with $d_{l/2}$ forms
a valid loop. This process is exponential in $l$, but as resonances
of size $l>6$ are extremely rare in the path integral as the highly correlated
structure which generates them is highly improbable, we restrict ourselves to $l=4,6$.
This avoids the exponential bottleneck and has no significant effect of efficiency
of sampling.\\

\noindent
4. Once a legitimate loop is identified, we fix the corresponding fuzzy region as
the upper end of our update column. Now we must identify the same loop at some slice
below $s_{init}$. We search sequentially the slices below $s_{init}$ until such a
region is found using the same technical apparatus discussed above. A successful
identification then defines the bottom end region.\\

\noindent
5. We now flip the column along the loop, i.e. the dimers go from living on $d_{l/2}$
to its complement on the loop. In the operator string language of SSEQMC, this involves
carrying all the operators living on the column to their new spatial positions as
defined by the complement. This operation may create some inconsistencies with the
path integral configuration which is not updated by this column flip. We check for
such inconsistencies at this stage, and if identified we abort the update. For the
type of resonances expected in a QSL, this is expected to happen rarely, and we see
that our update still succeeds with an $O(1)$ probability.\\

\noindent
6. To ensure that detailed balance is satisfied, we attempt to carry out the update
in reverse by beginning at the same initial slice $s_{init}$. If the same $d_{l/2}$ and unique
loop are found, then we carry out the flip along the loop in reverse. If at the end
of this operation, the path integral configuration is exactly the same as the one before
step 1 described above, then we have satisfied detailed balance, as probability
$P(init\to final)=P(final\to init)$ and $P(init)=P(final)$.
However, this is not always guaranteed and an example
of the same is shown in Fig.~\ref{FigQW}.
Applying the update on the final configuration
(shown on the right) leads to an identification of
a different column due to the arrangement of
creation/annihilation events. Thus the update
acting on the final configuration would not create
the initial configuration (left one), and we reject
this move as we cannot guarantee detailed balance.\\

The success of this update depends crucially on the presence of flippable columns.
A reliable proxy for the same is the density of flippable plaquettes.
We record this quantity in our simulations for the smallest
$2\times 2$ plaquette and show our results in Fig.~\ref{FigL4perf}c
for $\Omega=0.42$. In this regime, we expect a classical spin liquid
at intermediate temperatures and indeed find that the density
is lower than at low temperatures. At still lower temperatures, one
can see that the density saturates as the ground state has
been reached. Note that the temperature required for this saturation
is much lower
than all other scales in the problem, signifying that this is a low 
energy process.
We find that the success probability for this update also
saturates below a suitable temperature.
Although all the results discussed here are for $L=4$,
we expect the performance to be similar for larger system size,
as the success depends only on column size, which is a finite number
and is independent of $\beta$ (for large enough $\beta$)
and system size $L$.

\subsection{Test of accuracy of updates}

\begin{figure*}[t]
\includegraphics[width=\hsize]{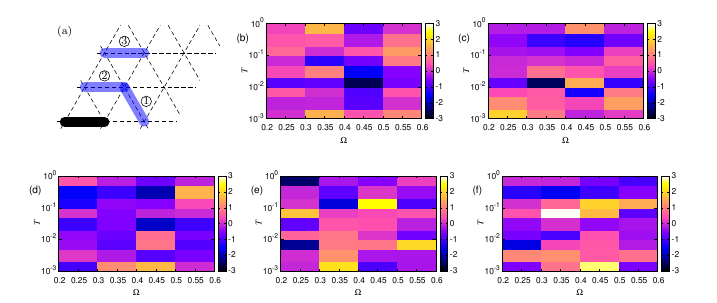}
\caption{
(a) Three correlation functions are calculated between the black bond,
and the three numbered blue bonds.
(d), (e) and (f) are the comparison with
ED (using $R/\sigma^N$ as defined in text)
for bonds number 1,2, and 3 respectively,
with a relative error bar $\sigma^N$ of $O(10^{-2})$, (b) is for the energy with
$O(10^{-4})$, and (c) for the dimer density with $O(10^{-3})$. In all cases,
we see an agreement within $\pm 3$, in units of the error bars.
}
\label{FigED}
\end{figure*}

To get an intuition about the behavior of our system and to check the
accuracy of our algorithm, we carry out a comparison with exact diagonalization
for a small system size. The triangular lattice can in general be written
as 3 lattice links (forming a triangle) sitting on each unit cell and
the unit cells are arranged in a parallelogram with periodic boundary
conditions. This is shown for $L=3$ in Fig.~\ref{FigTL},
and corresponds to a system with $3L^2$ links (Rydberg atoms).
For odd $L$, it is not possible to have a fully packed dimer covering,
as every dimer must have two unique end points. This implies that the
$L=3$ lattice must host one empty lattice site, and the number of occupied
links (excited Rydberg atoms) is at most four. The entire Rydberg allowed
Hilbert space corresponds to all possible hard core dimer coverings
(with any number of dimers). For $L=3$, the size of this space is $1126$,
and this can be diagonalized fully without difficulty. This allows us to
carry out a comparison with QMC at various values of temperature $T$
and $\Omega$.

To ensure that the QMC algorithm is sampling efficiently, we compare a
few different quantities against exact diagonalization (ED). The comparison
for an arbitrary observable $A$ is
quantified by calculating the relative difference between the QMC and ED
data, given by 
\begin{equation}
R=(A_{QMC}-A_{ED})/A_{ED},
\end{equation}
and normalizing $R$ by
$\sigma_{QMC}$, which is the error from the QMC calculation.
For all quantities we ensure that the normalized error
$\sigma^N=\sigma_{QMC}/A_{QMC}$ is $O(10^{-2})$ or smaller.
The exact values for different quantities are mentioned in the caption
of Fig.~\ref{FigED}.
If the deviation from ED is within the QMC error bars, we would expect
$R/\sigma^N$ to be $O(1)$, and this is the quantity plotted
in Fig.~\ref{FigED}.
The most relevant observable to check is the energy, and this is usually
the one with least noise in QMC. The comparison with ED is shown in
Fig.~\ref{FigED}, and we see that we can get at least a 4-digit match in this case.
Due to the stability of energy in QMC, it is not always considered a
good check of ergodicity, and as the updates we have proposed also
improve sampling of spatial structure, we would like to compare average
dimer density and various dimer-dimer correlations.
The comparison for the dimer density $\braket{n}$ is shown in Fig.~\ref{FigED},
and again we see a match which is of similar quality as the energy.
We probe spatial structure by considering dimer correlation functions
for three links with respect to a fixed link. These combinations are shown
in Fig.~\ref{FigED}.
Note that due to the translation symmetry of the lattice, we can calculate
the correlation function using all symmetry related pairs. However, here we
want to study the accuracy with which the QMC estimates the correlation
of a particular pair, and thus we do not use the symmetry related partners.
Even after this, we find an agreement within $1\%$ of the ED values,
for correlations which are themselves quite weak ($O(10^{-2})$). This is
a rigorous check that the QMC is able to efficiently sample the spatial
structure at all temperatures.

\subsection{Advantage of using the quantum resonance update}

To illustrate the utility of the quantum resonance, we study $L=4$ triangular
lattices at low temperatures for a value of $\Omega$
(specifically $\Omega=0.42$) which we expect is in
the correlated paramagnet regime.
This expectation is justified by our study of the specific heat and string
order parameters
later in this manuscript, which shows a
correlated paramagnetic behavior at
$\Omega_c\lesssim0.55$.
features at small scales.

\begin{figure}[t]
\includegraphics[width=\hsize]{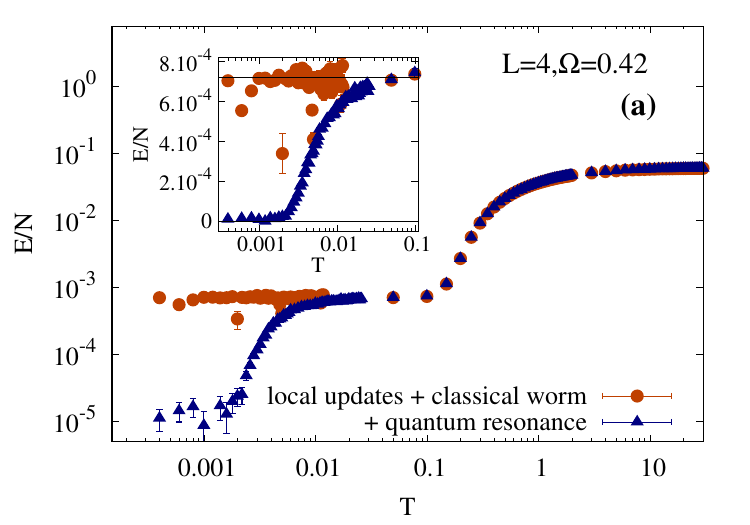}
\includegraphics[width=\hsize]{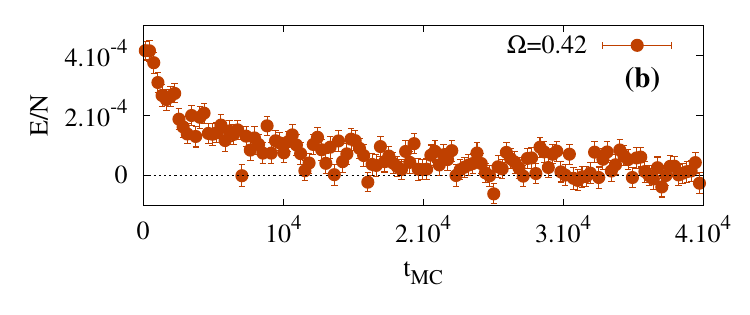}
\includegraphics[width=\hsize]{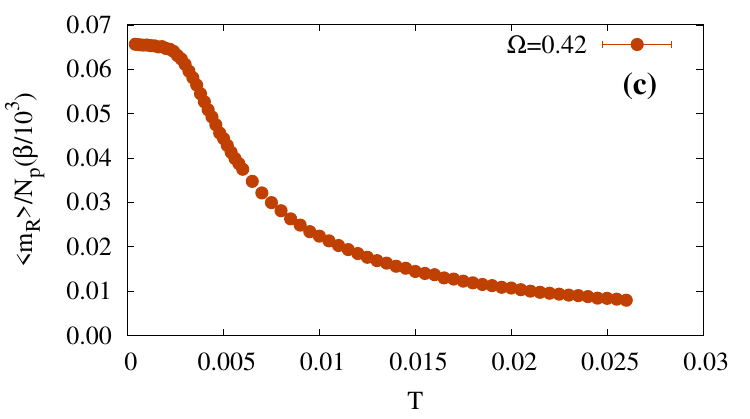}
\caption{
For $L=4$ at $\Omega=0.42$ :
(a) Energy per link as a function of temperature for different updates.
We shift the energy axis using our the ground state
energy from Lanczos diagonalization,
i.e. new ground state energy is zero. Without the quantum resonance
algorithm, the simulation is unable to sample the low $T$ regime,
inset zooms in on this using a linear scale for the energy.
(b) Reduction of energy as a function of Monte Carlo time once the
quantum resonance update is turned on at $T=0.001$.
(c)
Average number of resonances in the path integral, normalized by the
number of possible $2\times2$ resonating plaquettes and the average size
of imaginary time required to converge to the ground state ($10^3$).
}
\label{FigL4perf}
\end{figure}

At high temperatures, the system
samples all of Hilbert space uniformly, leading to a trivial paramagnet like
phase. Upon lowering temperature, we find a two step lowering of the
energy density,
first to the CSL, and then to the correlated paramagnet
(shown in inset of Fig.~\ref{FigL4perf}a). The zero is set here using the
ground state energy from exact diagonalization, thus showing that
the QMC approaches the correct value for $T\to 0$.
The lower reduction of energy density is
absent when we use only local updates and the
classical worm update, implying that the sampling method is
clearly insufficient.

Here we also show results for a more detailed study
of the ergodicity by first analyzing
the non-equilibrium behavior of our Monte Carlo simulation after
equilibrating using just the classical worm update and local
updates. We choose a temperature at which the ergodic simulations
(with the quantum resonance update) confirm convergence to the ground state,
and run our simulation without the quantum resonance update for $L=4$
in the correlated paramagnet
regime ($\Omega=0.42$). Once equilibrium is reached using
only the classical worm and local updates, the energy measured
converges to $-0.1738(1)$, which is significantly higher than the
true ground state energy $-0.174502003(1)$
from Lanczos diagonalization.
We call this as the zero of
our Monte Carlo time ($t_{MC}=0$), and now include the quantum resonance
update in our simulation. As shown in Fig.~\ref{FigL4perf}b,
the energy reduces to the required value in a fairly small number
of Monte Carlo steps, thus showing that the quantum resonance update
is able to access the true quantum ground state regime starting
from the non-equilibrium configuration generated by using only
the other updates.

To study the reason for the success of the quantum resonance update,
one can calculate the average number of resonances on flippable
plaquettes present in the path integral.
Note that here the exact density can be considered only qualitatively,
as we have calculated only the density of certain known and simple
motifs strictly consecutive in the path integral,
i.e. when the four operators required for a plaquette resonance
follow each other in imaginary time.
Although this is not likely to be true for most resonances
(as intermediate operators which lie on other lattice sites
are possibly present), this estimate is sufficient to give us
a sense of the behavior with reducing temperature.
We show our results in Fig~\ref{FigL4perf}c, and see that
the density of resonances (per unit space-time, i.e. normalized by $N\beta$) grows with reducing temperature
and saturates at the smallest temperatures (as expected when
we reach the ground state).

\begin{figure}[t]
\includegraphics[width=\hsize]{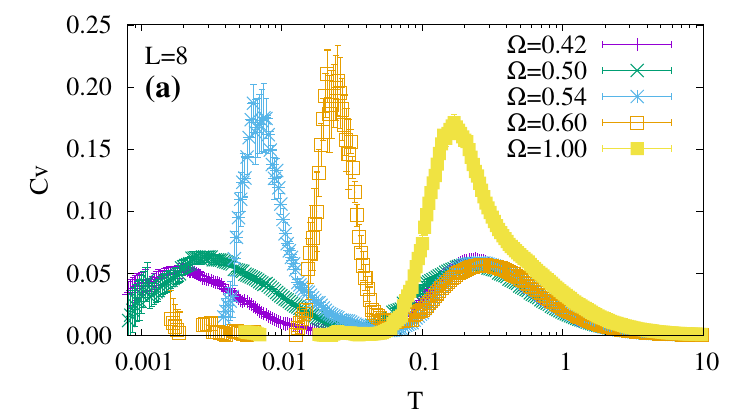}
\includegraphics[width=\hsize]{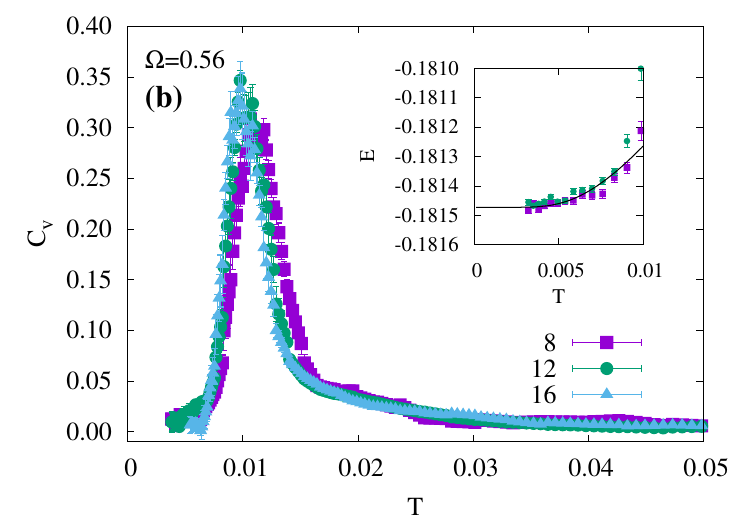}
\includegraphics[width=0.9\hsize]{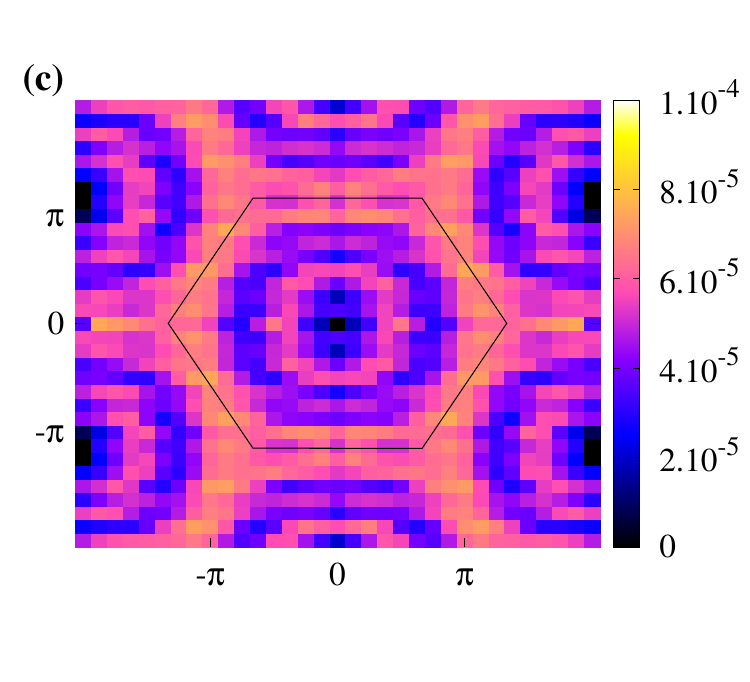}
\caption{
(a) Specific heat $C_v$ as a function of temperature $T$ for various values of
$\Omega$ for $L=8$. (b) $C_v$ at fixed $\Omega=0.50$ for a range of sizes
shows only weak finite size corrections. (c) Structure factor for $L=16$
at $T=0.0032$ and $\Omega=0.56$
}
\label{FigCvSF}
\end{figure}

\section{Results for Triangular lattice}\label{sRES}

Let us first discuss the possible phases realized on the triangular lattice
at low temperature.
For $\Omega\gg\delta$, we expect a simple paramagnet, with low dimer density,
and insignificant correlations between dimers. For $\Omega=0$, we should
expect a classical spin liquid, i.e. degenerate ground states corresponding to
fully packed dimer coverings. For $0<\Omega<\delta$, we can expect either
a spin liquid, or some ordering with a large unit cell. 
One of the most robust metrics to develop a phase diagram is the specific
heat, which we calculate from our Monte Carlo simulations by taking the
numerical derivative of the energy with respect to temperature. As
the temperatures we study are orders of magnitude smaller than the dominant
scale $\delta$, we find that this method has substantially smaller error
bars than those generated by
using the fluctuation-dissipation theorem\cite{sandvik2010computational}.

\begin{figure}[t]
\includegraphics[width=\hsize]{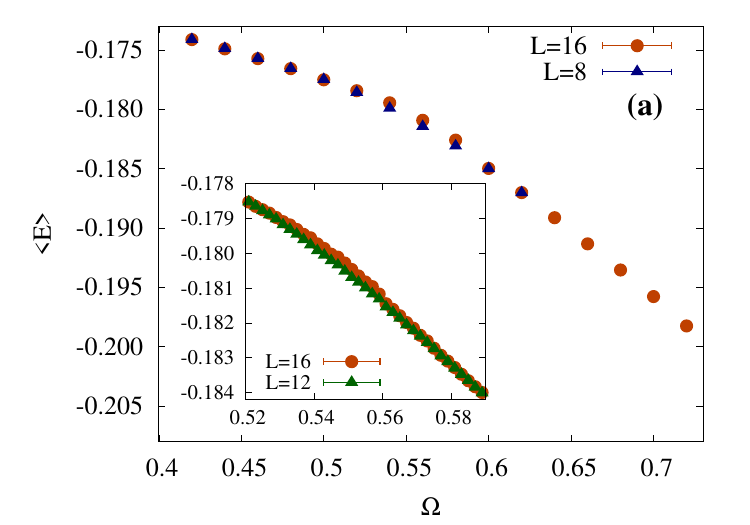}
\includegraphics[width=\hsize]{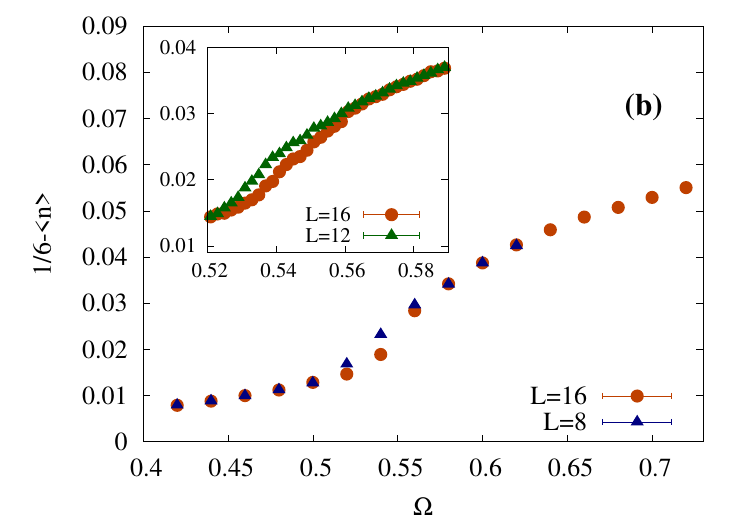}
\caption{
Energy (a) and monomer density (b) for a range of $\Omega$ at $T=0.0016$
and using $L=8$ and $16$ shows no sharp features, i.e. no quantum phase transition.
Zoomed in versions for $L=12$ and $16$ also shown for $T=0.0032$.
}
\label{FigEN}
\end{figure}

Our results are shown in Fig.~\ref{FigCvSF}a,
and we find that at $\Omega\approx\delta$, the specific heat shows a single
peak. In this regime we expect a quantum paramagnet for the ground state
and the specific heat peak is expected to be a smooth crossover from the
thermal paramagnet. Upon lowering $\Omega$, we find a double peak
structure, which signifies first the crossover to the classical spin liquid
regime when approaching from high temperature, and then a crossover
or transition to a unique ground state at low temperature. This is
consistent with the picture at $\Omega=0$, where we expect only a single
crossover from the high temperature paramagnet to the classical spin liquid.
Note that at $\Omega=0$, we expect fully packed dimer coverings at zero
temperature, as $\delta$ acts as a chemical potential for the dimers. However,
for $\Omega\neq0$, this is not the case, as the $\Omega$ term acts as a
creation(annihilation) operator and the system can lower its energy by having
a fluctuating number of dimers. Thus, in general, for all parameter regimes
with $\Omega>0$, we expect a non-zero density of monomers. To check that
we are indeed reaching the ground state for the parameter range for interest,
we integrate the specific heat to get the entropy as a function of temperature.
This analysis is shown in App.~\ref{AppA}, and we conclude that we are
reaching the zero entropy regime for the lowest temperatures that we are
able to access.

\begin{figure}[t]
\includegraphics[width=\hsize]{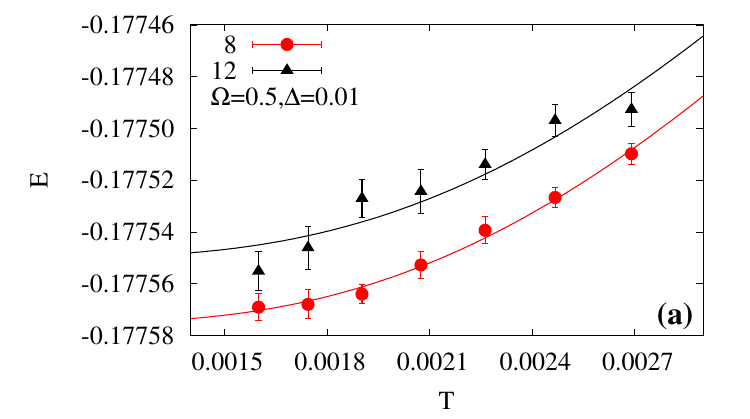}
\includegraphics[width=\hsize]{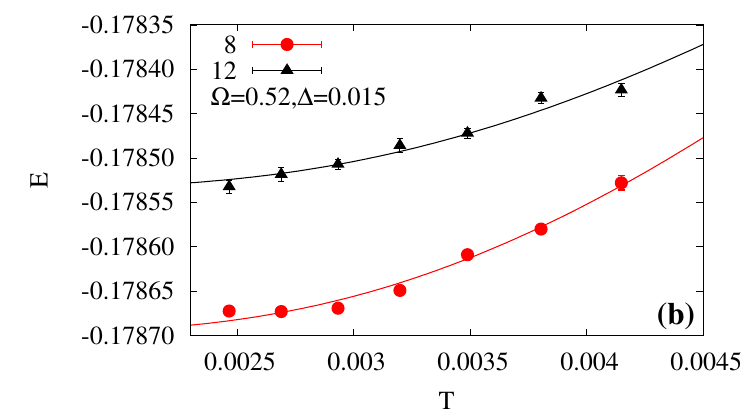}
\includegraphics[width=\hsize]{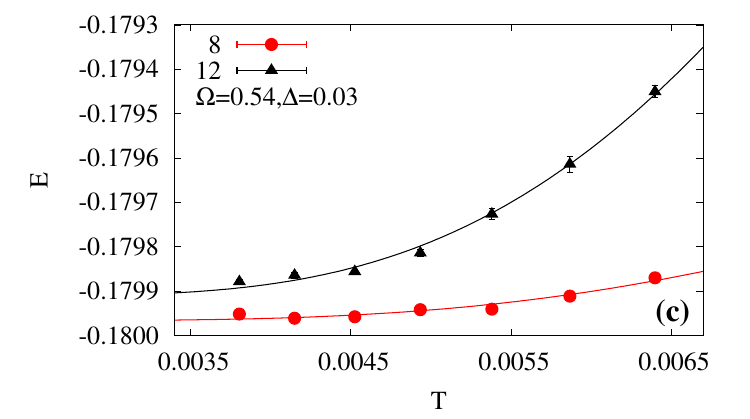}
\caption{
Energy as a function of temperature for
$\Omega=(a)0.5, (b)0.52$ and $(c)0.54$. Fits to $a+be^{-\Delta/T}$
with the same $\Delta$ for system size $L=8$ and $12$.
The fitted value of $\Delta$ is noted in the legend
of each plot.
}
\label{FigGap}
\end{figure}

\begin{figure*}[t]
\includegraphics[width=\hsize]{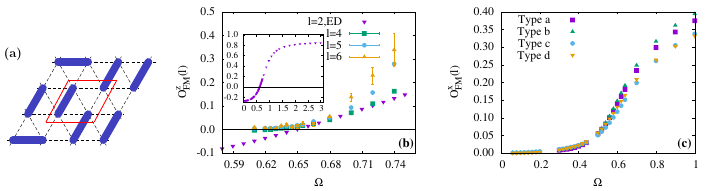}
\caption{
(a) Contour for $O^z_{FM}$.
(b) Vanishing behavior for $L=12,T=0.0128$ as we move away from strongly
paramagnetic regime.
(c) Behavior consistent with $O^x_{FM}$ for $L=4$ exact diagonalization
for ground state on loop configurations $a$-$d$ described in App.~\ref{AppB}.
}
\label{FigFM}
\end{figure*}

\subsection{Investigation of quantum phase transition}

The monomer density 
($1/6-$dimer density)
and energy as a function of $\Omega$
are shown in  Fig.~\ref{FigEN}.
There is a rapid change in the behavior of both quantities
close to $\Omega=0.56$, which
may lead one to suspect a quantum
phase transition (QPT) from 
$Z_2$ QSL to trivial paramagnet around this value of $\Omega$.
However, as we discuss below, analysis
of the system size dependence and thermodynamic properties of our simulations
does not support this conclusion.
Instead, the simulations seem to show a smooth
crossover in this regime, leading us to the
conclusion that the putative QSL phase is not truly distinct from the trivial quantum
paramagnet.

The inflection points in the
energy and monomer density in Fig. 
\ref{FigEN} do not
get significantly sharper on increasing system size from $L=8$ to $L=16$, which
indicates against a QPT.

The absence of a quantum phase transition is further indicated from our $C_v$ data
(Fig.~\ref{FigCvSF}a), which shows the gap (the lower temperature peak
can be considered to be a proxy for this) is smoothly reducing with
$\Omega$, rather than closing and opening again as we would expect for a
transition between two gapped phases.

In the inset of Fig.~\ref{FigCvSF}b, we have
plotted the low temperature behavior of the energy
per link, and fit to the form $a+be^{-\Delta/T}$,
with the same value for the gap $\Delta$ for system
size $L=8$ and $12$. We find a good fit for 
$\Delta=0.03$ for $\Omega=0.56$. We also
check the range of $\Omega\in(0.5,0.54)$ and
find a finite gap for this region which is reducing
with $\Omega$ as expected. These fits are shown in
Fig.~\ref{FigGap} and leads us to conclude that
there is no closing of the gap at least down to
$\Omega=0.5$.

To confirm that this is indeed a gapped point and
not a QPT, we study $C_v$ for $L\in\{8,16\}$ and find no significant finite
size effects, and an exponential decay of energy as $T\to0$ (Fig.~\ref{FigCvSF}b).
This signifies a finite energy gap at $\Omega=0.56$.
The $C_v$ data shown in 
Fig.~\ref{FigCvSF}b
clearly indicates that the two peaks seen in the temperature scans are
crossovers and do not sharpen into phase transitions with increasing size.
This confirms our expectation that the ground state is paramagnetic, and
as a secondary check we can study the structure factor.
Connecting the link centers of the triangular
lattice leads to a Kagom\'e lattice, and thus
we can use the definition of the structure factor for a spin system
living on the same.
Following
Ref.~\cite{sherman2018structure},
we define the structure factor as
\be\label{eqSF}
S(\vec{q})=\sum_{ab}\sum_{i_1,i_2}
\langle n_a(0,0)n_b(i_1,i_2)\rangle
e^{-i\vec{q}\cdot(i_1\vec{R}_1+i_2\vec{R}_2
+\vec{l}_b-\vec{l}_a)},
\ee
where $(i_1,i_2)$ indexes a unit cell of the 
Kagom\'e lattice, $(a,b)$ index the site numbers within that unit cell
and the lattice vectors defined as 
$\vec{R_1}=(2,0)\ ,\ \vec{R_2}=(1,\sqrt{3})\ ,\ \vec{l_0}=(0,0)\ ,\
\vec{l_1}=(-\frac{1}{2},\frac{\sqrt{3}}{2})\ ,\
\vec{l_2}=(\frac{1}{2},\frac{\sqrt{3}}{2})$ based on the geometry
shown in Fig.~\ref{FigTL}.
The correlation function $\langle n_a(0,0)n_b(i_1,i_2)\rangle$
used here is the connected version with $\langle n\rangle^2$ 
subtracted, where $\langle n\rangle$ is the mean dimer density.
For $L=16,\Omega=0.5$ and $T=0.0032$, 
as shown in Fig.~\ref{FigCvSF}c, we find a
featureless $S(\vec{q})$, supporting our previous
conclusion that the ground state is either a paramagnet or a $Z_2$ QSL.
In contrast to dimer models on bipartite lattices, it is known
\cite{fendley2002classical}
that for the triangular lattice, even a fully packed classical dimer model
has extremely short range correlations. This implies that no pinch point
like structures are expected in $S(\vec{q})$.

It may be argued that since
our simulations are conducted at finite
temperature, a smooth crossover from a $Z_2$ QSL to a trivial paramagnet is expected in two dimensions, and that our results therefore
do not contradict the existence of a $Z_2$ QSL at $T=0$. However, we note that our simulations reach the zero entropy regime, and so we would expect to see signatures of any singular behavior at $T=0$ in our simulations. Moreover, even at finite temperature we would expect to see the effects of a gap closing in the thermodynamic properties in the vicinity of a QPT, which is not found in our simulations.

\subsection{String and Fredenhagen-Marcu order parameters}
Dimer coverings with a small number of monomers can be expected to retain
the essential physics of fully-packed coverings. The topological character of dimer coverings is typically
identified using string order parameters. However, even an arbitrarily
small monomer density leads to string order parameters vanishing in the
thermodynamic limit. This has lead to the use of the Fredenhagen-Marcu
order parameter (FMOP) \cite{FMorg}, which can robustly identify topological
signatures in such regimes. This has recently been used to characterize
a topological transition in the perturbed toric code \cite{FP},
and to identify a possible $Z_2$ spin liquid in a Rydberg atom array
on the ruby lattice\cite{Vryd}. Following the latter study, we define
\be\label{efm}
O^{\alpha}_{FM}(l)=\frac
{\braket{\prod_{i\in o_l}\sigma^{\alpha}_i}}
{|\braket{\prod_{i\in c_l}\sigma^{\alpha}_i}|^{\frac{1}{2}}} 
\ee
where $o_l(c_l)$ corresponds to a open(closed) contour of size $l\cross l$
(shown in Fig.~\ref{FigFM}),
and $\alpha=x,z$ correspond to off-diagonal and diagonal terms respectively.
The numerator and denominator individually
are the open and closed string order parameters respectively.
In terms of the creation/annihilation and density operators discussed above,
$\sigma^x_i=b^{\dagger}_i+b_i$ and $\sigma^z_i=2n_i-1$.
For a superposition of fully packed dimer coverings,
the string order parameter along a closed
contour (denominator in Eq.~\eqref{efm})
is unity irrespective of its perimeter. However, a vanishing density
of dynamical monomers lead to an exponential decay with perimeter for the
same quantity. In this case, the FMOP is the ratio of two exponentially
decaying quantities, and acts as a proxy for the weights of open-string quantum
fluctuations (movement of a single monomer) and of
closed-string quantum fluctuations (flip of dimers along a
closed contour)\cite{FP}.

\begin{figure}[t]
\includegraphics[width=\hsize]{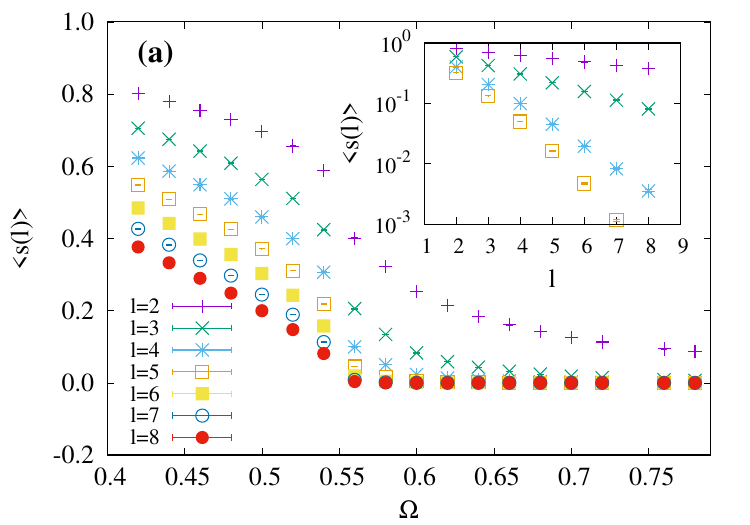}
\includegraphics[width=\hsize]{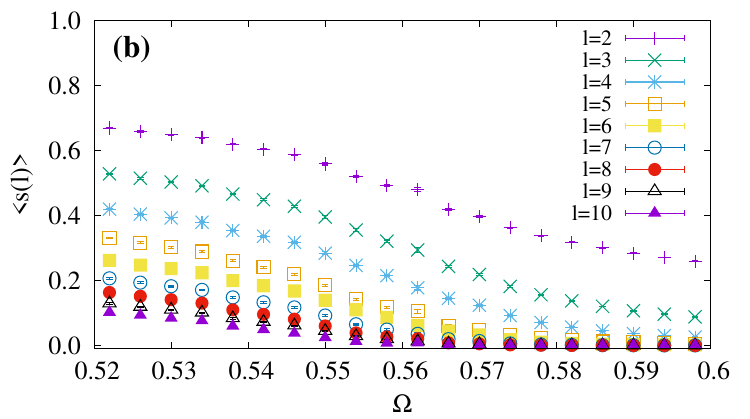}
\caption{
String order parameter defined over the contour shown in Fig.~\ref{FigFM}a 
with $l=2$
for a $4\times 4$ system.
(a) $\braket{s(l)}$ for $L=16$ at $T=0.0016$ for various $l$ shows an
increasing trend with reducing $\Omega$. Inset: Extrapolation of $\braket{s(l)}$
at $\Omega=0.42,0.54,0.56$ and
$0.58$ as a function of $l$
(Faster decays with increasing $\Omega$). The last three data points (largest
three values of $l$) are fit to $\braket{s(l)}=ae^{-l/b}$ to be consistent with
$\lim_{l\to\infty}\braket{s(l)}\to 0$.
(b) Same for $L=20$ close to $\Omega=0.56$ for $T=0.0045$, which is low
enough to capture ground state physics for this range of $\Omega$ shows no
sharp behavior.
}
\label{FigStr}
\end{figure}

For a $Z_2$ spin liquid, both $O_{FM}^x$ and $O_{FM}^z$ should vanish for large
$l$\cite{Vryd}. In contrast, for the paramagnetic regime,
$\Omega\approx\delta$, the numerator and denominator of Eq.~\eqref{efm}
both decay exponentially in a way which leaves the FMOPs finite. However,
due to extremely small numerical values in the ratio, it is difficult to
get small relative error bars for either $O_{FM}$ from Monte Carlo simulations.
An added complexity for $O_{FM}^x$ is the fact that it is composed of
strings of off-diagonal operators which suffer from extreme statistics in
the SSEQMC formulation\cite{SandvikReview}. Thus we are only able to extract
$O_{FM}^z$ up to $l=6$ with reasonable statistics from our simulations.
To study $O_{FM}^x$, we resort to exact diagonalization for a $L=4$ system
with various shapes for the contour. We fix temperature $T=0.0032$, which is
expected to be low enough for the range of $\Omega$ we consider based on
the $C_v$ data presented in Fig.~\ref{FigCvSF},
and present our results for the FMOPs in Fig.~\ref{FigFM}.
We see that $O_{FM}^z$ reduces smoothly with decreasing $\Omega$, indicating
that closed string dynamics are dominant over open string. This data is
collected for $L=12$ at $T=0.0128$, which corresponds to the ground state
for $\Omega>0.6$. The various loop geometries we have used for $O_{FM}^x$
are listed in App.~\ref{AppB}, and we see in Fig.~\ref{FigFM}c
that all of them show a similar growth from zero close to $\Omega=0.5$.
Thus we have signatures consistent with a $Z_2$ spin liquid for
$\Omega \lesssim 0.5$.

We detect the same in the string order parameter as well,
on a closed contour for $L=16$, as  shown in Fig.~\ref{FigStr}
for various loop sizes $l$. Although its value is expected to vanish for
$l\to\infty$ (inset of Fig.~\ref{FigStr}), we see a significant growth even for
$l=8$ (which is the largest possible for $L=16$), with a change in behavior
around $\Omega=0.56$.

Thus, although we have argued
above on the basis of thermodynamic evidence
that we do not see a QPT in this model, 
the FM order parameters do show signatures
consistent with a $Z_2$ QSL for $\Omega \lesssim 0.56 $. To resolve this, we propose
that there is a regime for $\Omega \lesssim 0.56$ where the behavior of a  $Z_2$ QSL is
observable up to a finite length scale. This
length scale is tunable with $\Omega$ 
and becomes smaller than a single lattice spacing for $\Omega > 0.56 $.


\subsection{Tunable scale for implementation of dimer constraint}

\begin{figure}[h]
\includegraphics[width=\hsize]{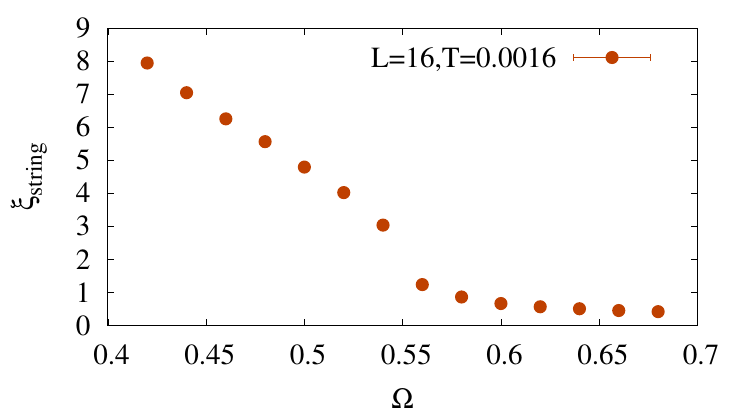}
\caption{
Length scale for topological protection
$\xi_{string}$ as a function of $\Omega$
for $L=16,T=0.0016$, shows a smooth decay
with increasing $\Omega$.
}
\label{FigTS}
\end{figure}

Our results for the string order parameter on a closed loop
suggest an exponential decay with loop size at a fixed $\Omega$.
This is shown in the inset of Fig.~\ref{FigStr}a,
and allows us to define a length scale over which we can expect
the topological protection engineered by the fully packed dimer constraint.
The closed string order parameter for the $l\times l$ loop can be considered
to decay as $ce^{-l/\xi_{string}}$ and we can fit the numerical data to extract
$\xi$ as a function of $\Omega$. The result of the fitting process is shown
in Fig.~\ref{FigTS} and we see that this scale smoothly decreases with
increasing $\Omega$. The scale $\xi$ can thus be considered to be a proxy for
the degree of topological protection, i.e. sizes $<\xi_{string}$ will behave
as if they are perfectly topological and those with sizes $>\xi_{string}$  will
behave like simple quantum paramagnets.

The PXP model we have studied results in a quantum dimer model (QDM) at
fourth order in perturbation theory which has only a kinetic term,
i.e. $V=0$ in the QDM language \cite{moessner2001resonating}.
It is known that for the triangular lattice QDM, this parameter range hosts
an ordered ($\sqrt{12}\times\sqrt{12}$) ground state\cite{moessner2001resonating}.
This supports our finding that there is no true $Z_2$ spin liquid for
the small to intermediate values of $\Omega$ we have studied, 
although we do not see 
translational symmetry breaking in our simulations.
%
For intermediate values of $\Omega$
we see a regime that seems smoothly connected to a quantum paramagnet, albeit with
strong remnant similarities to a $Z_2$
spin liquid for intermediate scales.
At finite temperature, the crossover from the CSL to
the quantum paramagnet when increasing $\Omega$
can look like a phase transition due to a quick
change in dimer density and the string order
parameter. This is discussed in detail in
App.~\ref{AppC} for $T=0.0128$ (which is much
smaller than the microscopic energy scales),
and implies that care must be
taken when trying to identify a quantum phase
transition from numerical simulations
\cite{yan2022triangular}.

\section{Conclusions}

We have studied the PXP model in the reduced Hilbert space enforced
by Rydberg blockade constraints on a triangular lattice. Our numerical
simulations show the presence of a paramagnetic ground state for the
relevant range of the strength of laser driving $\Omega$,
with a tunable scale for topological protection
built in and controlled by $\Omega$.
We arrive at the conclusion about topological protection using an
appropriately defined string order parameter, and studying its
behavior with varying $\Omega$.
A similar behavior of the string order parameters has been seen for
the Kagom\'e lattice\cite{wang2024renormalized} where the QMC simulations
have been unable to access the quantum spin liquid regime, but
have found similar partial topological protection in the
classical spin liquid regime.
Our results suggest
that topological protection up to a tunable length scale
can be engineered in Rydberg atom arrays, and that this
differs from the perfect topological protection seen in quantum dimer
models as it hosts dimer
coverings which are not fully packed. This may even be a possible interpretation
for recent observations of topological protection in a triangular lattice
quantum dimer model with monomers \cite{yan2022triangular}, where
topological features at finite scale (in particular the string order
parameter for a small loop)
have been interpreted
as evidence for a $Z_2$ QSL.
A recent study for a Rydberg atom array
in 3D has also found similar results
\cite{wang2025doped}, i.e.
monomers may lead to a trivial state
but for small system size spin liquid
signatures (in their case a $U(1)$
Coulomb liquid is considered)
can be retained.


To achieve the results we have presented in this paper, we have
introduced a sampling of quantum resonances in the path integral
representation. We have shown
that without this sampling method,
the crossover between the CSL and the quantum paramagnetic regimes
of interest is impossible to
capture, thus making it a necessary ingredient in our numerical
recipe. This is especially important to Rydberg atom arrays, as
this is a notoriously hard regime to sample using Monte Carlo
simulations, with hurdles having been recently identified
\cite{yan2023emergent,hibat2024recurrent}.
As the resonances sampled are expected to be a general
feature at least for $Z_2$ QSLs and similar phases \cite{zhang2024quantum}, our
algorithm can readily be adapted to other model Hamiltonians which may host such states,
such as quantum dimer models, XXZ models on the pyrochlore lattice,
and frustrated transverse field Ising models.

\begin{figure}[t]
\includegraphics[width=\hsize]{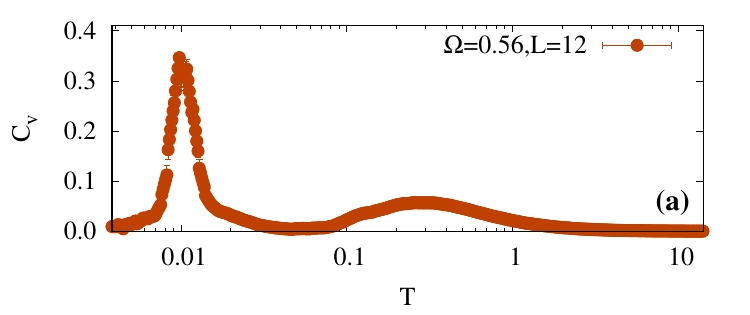}
\includegraphics[width=\hsize]{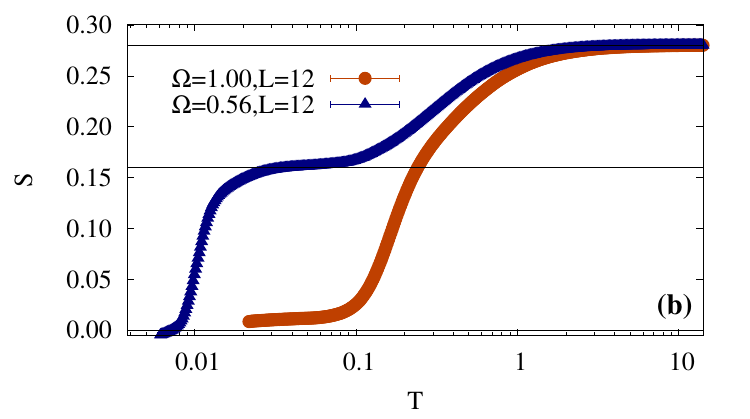}
\caption{
(a) Specific heat for $L=12$ as a function of temperature shows two
well defined peaks.
(b) Integration of $C_v/T$ for $\Omega=1$ and $0.56$ shows the
residual entropy plateaus corresponding to the CSL and the full hard
core dimer Hilbert space.
}
\label{FigL12}
\end{figure}

\begin{figure*}[t]
\includegraphics[width=\hsize]{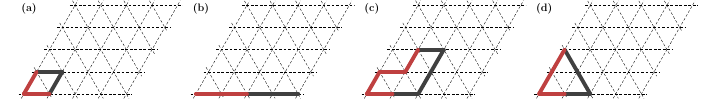}
\caption{
The various patterns of closed loops used to calculate $O^x_{FM}$ in Fig.~\ref{FigFM}.
Red bonds corresponds to the portions used to calculate the open segments.
}
\label{FigFMp}
\end{figure*}

\section{Acknowledgements}
We would like to thank Zhenjiu Wang, Frank Pollmann,
Claudio Castelnovo, Ciaran Hickey, Rhine Samajdar,
Han Yan, Zheng Yan and Lode Pollet for useful discussions.
Numerical simulations for this project were carried out
using resources provided by the Max-Planck-Institut
f\"ur Physik komplexer Systeme (MPIPKS)
and Okinawa Institute of Science and Technology.
We also acknowledge the support of the MPIPKS, where this project began.

\appendix
\section{Specific heat and entropy for $L=12$}\label{AppA}

To illustrate the utility of the quantum resonance, we study $L=12$ triangular
lattices at low temperatures for a value of $\Omega$
(specifically $\Omega=0.56$) which we expect is in
the correlated paramagnet regime.
As discussed in the main text, the specific heat should show two peaks
as a function of temperature in this regime.
This is shown clearly in
Fig.~\ref{FigL12} for $L=12$, where we have used both the classical worm and quantum
resonance updates, and local updates developed in Ref.~\cite{patil2023quantum}.
As we calculate the specific heat by calculating the derivative of energy
versus temperature, and cannot achieve small error bars for low temperatures
due to the large computational resource requirement, $C_v$ at
temperatures below $0.005$ have substantial error bars.
However, at these temperatures the simulation is already sampling
the ground state, and thus not relevant to our study.

The specific heat data for $L=12$ can be utilized to study the
thermodynamic entropy as a function of temperature. This is done by
integrating $C_v/T$, and usually requires one to know the value of
the entropy at one of the limits (either $T\to0$ or $T\to\infty$).
As the restricted Hilbert space in which our model operates has
an unknown entropy at $T\to\infty$ (as the states are made of
complex dimer coverings), we must use the opposite limit,
i.e. $T\to0$.
This limit can be utilized in the paramagnetic regime
($\Omega\approx 1$), where we know
that the ground state should have zero residual entropy. We
use the integral at $\Omega=1$ to get the entropy of the restricted
Hilbert space as $T\to\infty$, and find this to be $0.28(1)$
per link, which is significantly smaller than $ln(2)\approx 0.693$, which
would be the case for uncorrelated links. The entropy as $T\to\infty$
should be the same for all values of $\Omega$, and thus we use this
value as the upper limit when performing the integration in the
correlated paramagnet
regime at $\Omega=0.56$. We find that there is a plateau at
intermediate temperatures which is associated with the CSL, with
a residual entropy per link of $0.16(1)$, and an eventual drop to zero entropy
within the error bars from our specific heat data. This is shown for our
$L=12$ $C_v$ data in Fig.~\ref{FigL12}, and leads to the conclusion that in the
correlated paramagnet regime we are able to reach temperatures where we have converged to
the ground state.

\section{Loop arrangements for calculating $O^x_{FM}(l)$ from exact
diagonalization}\label{AppB}

As mentioned in Sec.~\ref{sRES}, the off-diagonal FM order parameter
$O^x_{FM}(l)$ cannot be calculated to high accuracy using QMC due
to large fluctuations in the estimators. Thus the data presented for
the same in Fig.~\ref{FigFM}c
is for an $L=4$ lattice using exact diagonalization. As this prevents
us from performing any scaling with loop size, we instead show 
in Fig.~\ref{FigFM}c that
the behavior of $O^x_{FM}$ is robust for different loop shapes.
The loop configurations for which we have calculated this data
(denoted by type $a$-$d$ in Fig.~\ref{FigFM}c)
are shown in Fig.~\ref{FigFMp}.
Each loop is of even length, and the open string is highlighted in red,
whereas its partner which completes the closed loop is shown in black.

\begin{figure}[t]
\includegraphics[width=\hsize]{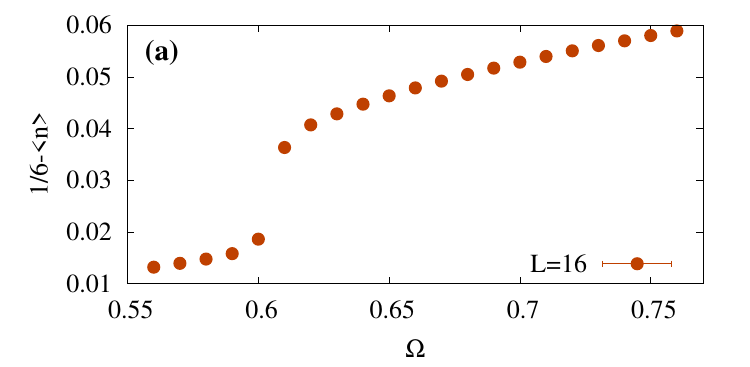}
\includegraphics[width=\hsize]{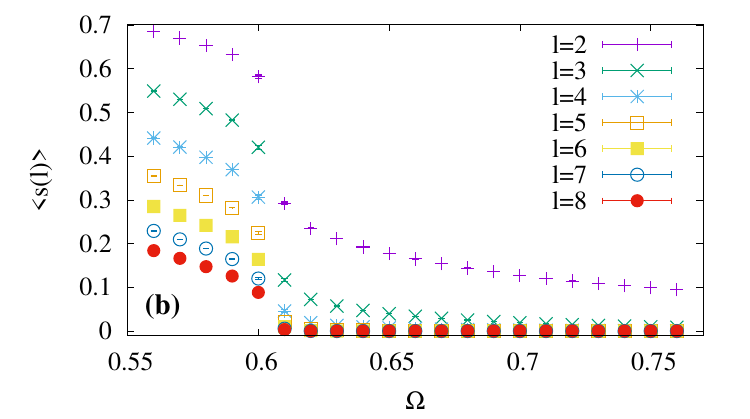}
\caption{
(a) Monomer density and
(b) string order parameter for closed strings
for a range of $l$ as a function of $\Omega$ for
$T=0.0128$ and $L=16$.
}
\label{FigClCrs}
\end{figure}

The single plaquette resonance loop in Fig.~\ref{FigFMp}a
can be taken to be a direct estimate of the smallest quantum resonance
possible on the triangular lattice. Fig.~\ref{FigFMp}b
represents a loop which winds around the system, and is thus topologically
non-trivial. However, as seen in Fig.~\ref{FigFM}c,
this does not affect the behavior as we are not considering a
fully packed dimer model. Fig.~\ref{FigFMp}c represents the longest
loop which we have tested, and is made up of eight links and does
not enclose any sites of the triangular lattice.
Fig.~\ref{FigFMp}d
is the only loop which we have considered where the open string has
an odd number of links. However, we find that this does not significantly
affect its behavior as a function of $\Omega$.

\section{Crossover between classical spin liquid and quantum paramagnet}
\label{AppC}

Here we discuss the crossover between the classical
spin liquid regime and the quantum paramagnet when
tuning $\Omega$ at finite $T$. We choose $T=0.0128$
as we have seen in Fig.~\ref{FigCvSF} that this
temperature is higher than the location of
the low temperature peak. This implies that
there is still some residual entropy at this
temperature for $\Omega=0.56$, which is close to
where we see the crossover. As shown in
Fig.~\ref{FigClCrs}, we see that the monomer density
and the string order parameter (as defined in
Sec.~\ref{sRES}) rapidly change near the crossover.
The CSL regime is characterized by the monomer density
vanishing, and the
string order parameter being sizable for small $l$.
Note that as the monomer density is non-zero in the range of $\Omega$
which we have studied.
This implies that for $l\to\infty$, the string order
parameter should vanish, and this is consistent with
the data presented in Fig.~\ref{FigClCrs}.

\bibliography{scpreferences.bib}

\end{document}